\documentclass[structabstract]{aa}  
\usepackage{graphicx}
\usepackage{natbib}
\usepackage{txfonts}
\begin{document}
\def\frac{$''$\hspace*{-.1cm}}
\def\deg{$^{\circ}$}
\def\min{$'$}
\def\deg{$^{\circ}$\hspace*{-.1cm}}
\def\min{$'$\hspace*{-.1cm}}
\def\h2{H\,{\sc ii}}
\def\hi{H\,{\sc i}}
\def\hb{H$\beta$}
\def\ha{H$\alpha$}
\def\sm{$M_{\odot}$}
\def\ab{$\sim$}
\def\x{$\times$}
\def\sec{s$^{-1}$}

   \title{{\it Spitzer} Mid-infrared Study of Compact \h2 Regions in the
     Magellanic Clouds}


   \author{V. Charmandaris
          \inst{1,2,3}
          \and
          M. Heydari-Malayeri\inst{3}
          \and
          E. Chatzopoulos\inst{1}
          }

   \institute{University of Crete, Department of Physics, Heraklion,
     GR-71003, Greece\\
              \email{vassilis@physics.uoc.gr}
              \and
             IESL/Foundation for Research and Technology - Hellas,
  GR-71110, Heraklion, Greece
         \and
             Observatoire de Paris, LERMA, 61 Avenue de l'Observatoire,
             F-75014, Paris, France
             }

\authorrunning{Charmandaris et al.}
\titlerunning{Compact \h2 Regions in the mid-IR}

\date{Received February 27, 2008; accepted May 16, 2008}

 
  \abstract
   {}
   { We present a study of the mid-infrared properties and dust
     content of a sample of 27 \h2 ``blobs'', a rare class of
       compact \h2 regions in the Magellanic Clouds. A unique feature
       of this sample is that even though these \h2 regions are of
       high and low excitation they have nearly the same physical
       sizes $\sim$1.5--3 pc.}
   {We base our analysis on archival 3 -- 8 $\mu$m infrared imagery
     obtained with the Infrared Array Camera (IRAC) on board the
     Spitzer Space Telescope.}
   {We find that despite their youth, sub-solar metallicity and varied
     degrees of excitation, the mid-infrared colors of these regions
     are similar to those of typical \h2 
regions. Higher excitation
     ``blobs'' (HEBs) display stronger 8 $\mu$m emission and redder
     colors than their low-excitation counterparts (LEBs). }
   {}


   \maketitle
%

\section{Introduction}

The detailed study of massive star formation in galaxies has been a
challenging research area over the years. Two of the main reasons are
that massive stars are intrinsically rare and their lifetime is short
\citep[see][for a review]{Zinnecker07}. In addition they are
predominantly formed not in isolation but in tight groups in dense
molecular clouds, where they are enshrouded by large quantities of
dust. As a result, probing massive stars at their early stages of
evolution is particularly difficult, since they are not readily
observed in the ultraviolet and optical wavelengths which give direct
access to their physical parameters. At this stage they can only be
detected indirectly by their infrared and radio emission, which is due
to emission from the surrounding dust and the associated ionized \h2
region respectively. As massive stars evolve though, their far-UV
photons and strong winds dissociate the molecules in the surrounding
clouds and ionize the atoms creating ultra-compact \h2
regions. Eventually, the natal molecular cloud is fully ionized to
become a compact \h2 region. As the front expands and the volume of
the ionized gas increases, the advancing ionization front of the \h2
region reaches the outer surface of the molecular cloud. Then the
ionized gas flows away into the interstellar medium according to the
so-called champagne effect \citep[]{Tenorio79,Bodenheimer79} and
consequently the newborn stars become accessible to observation in the
ultraviolet and visible.

A number of very young, emerging \h2 regions were identified in the
Magellanic Clouds on the basis of ground-based observations at the
European Southern Observatory, nearly 20 years ago
\citep[][]{Heydari82,Heydari83,Heydari85,Heydari86,Heydari90,
  Testor85,Heydari88a}. Searching for this type of regions in the
Magellanic Clouds was motivated by the fact that the interstellar
extinction along this line of sight is lower than towards the disk of
our Galaxy and that the lower metallicity of the Small and Large
Magellanic Clouds (SMC and LMC respectively) favors the formation of
massive stars \citep{Wolfire87,Melena08}.

This distinct and very rare class of \h2 regions in the Magellanic
Clouds was called \h2 ``blobs'' since no features could be
distinguished with the available telescopes at the time. They were
classified as high-excitation and low-excitation compact \h2 ``blobs''
(HEBs and LEBs respectively), principally based on the temperature and
mass of their exciting stars, as well as their nebular \hb\ luminosity
\citep[for details see][]{Meynadier07}. Contrary to the typical \h2
regions of the Magellanic Clouds, which are extended structures with
sizes of several arc minutes corresponding to physical scales of more
than 50\,pc and powered by a large number of exciting stars, HEBs and
LEBs are dense and small regions of $\sim$\,5\frac\, to 10\frac\, in
diameter in the optical, corresponding to $\sim$\,1.5--3.0\,pc.
Studies in the optical reveal that they are heavily affected by local
dust \citep[][]{Heydari88a,Israel91}.  A noteworthy characteristic of
the HEBs is that they harbor the youngest massive stars accessible to
infrared and optical observations. Since for massive stars the
accretion time-scale is larger than the Kelvin-Helmholtz time-scale,
they reach the main sequence while accretion is still taking
place. Moreover, as pointed out before, massive stars evolve very fast
and as a result obtaining their physical parameters ``at birth'' is a
fairly challenging task!  It is now believed that the ``blobs''
correspond to the final stages in the evolution of the ultra-compact
\h2 regions, whose Galactic counterparts are detected only at infrared
and radio frequencies \citep[][]{Churchwell90}. The study of HEBs thus
bridges a gap between understanding the properties of completely dust
enshrouded stars inside ultra-compact \h2 regions and those of the
exciting stars in evolved \h2 regions \citep[][]{Heydari07}.  It
should be noted that an important characteristic of our compact \h2
region sample is their uniform physical size which varies by less than
a factor of two among all ``blobs''. This additional constraint makes
them an ideal laboratory to study problems related to the formation of
massive stars and their environment within a well defined region.
Because of the contamination by strong nebular background no direct
information about the number of the exciting stars of the blobs was
possible with ground-based telescopes.  A detailed high spatial
resolution imaging and spectroscopy campaign of a number of those
regions was performed with the Hubble Space Telescope and several of
those issues were addressed \citep[see ][ and references
therein]{Heydari99,Heydari03,Heydari07}.

The complex interaction between the ionizing radiation of the young
stars, their dusty cocoons and adjacent molecular clouds can be best
studied in mid-infrared (mid-IR) wavelengths. At this wavelength range, not
only we are less affected by extinction, but also a number of
ionization lines (such as [S\,{\sc{iv}}] 10.51 $\mu$m,
[Ne\,{\sc{ii}}] 12.81 $\mu$m, [Ne\,{\sc{iii}}] 15.55 $\mu$m,) which
probe the chemistry the interstellar gas and radiation field are
available. Furthermore, the strength of broad spectral features, such
as those emitted by Polycycyclic Aromatic Hydrocarbons (PAHs) are
readily accessible and these can be used to trace the properties of
the surrounding dust and molecules as well as estimate the star
formation rate \citep[see][and references therein]{Peeters02,Forster04,
  vanDishoeck04,Calzetti07}.

A number of authors have already explored the mid-IR properties
of \h2 regions in the Magellanic Clouds with the Infrared Space
Observatory (ISO) and with the Spitzer Space Telescope
\citep[i.e.][]{Vermeij02,Jones05,Gouliermis07,Lebouteiller07}. None of
those studies though examined the compact \h2 region sample as a whole
with sufficient spatial resolution. The more detailed spectroscopic
study of \citet{Vermeij02} used ISO/PHOT-S to explore the variation of
the PAH features in the 5.8 to 11.6 $\mu$m range. This work did reveal
some interesting correlations between the strength of the PAH
features, as well as a difference in the total PAH emission normalized
by the strength of the radiation field between \h2 regions in the
Magellanic Clouds and those in our Galaxy. However, the ISO aperture
used was rather large ($24\arcsec\times24\arcsec$) and the spectra
clearly included not only the emission from the ``blobs'' but also
from adjacent star forming and photo-dissociation regions.

Here we present for the first time a mid-IR high resolution study of
the HEBs and LEBs in the 3.6 to 8.0$\mu$m range based on archival
Spitzer Space Telescope imagery. This is a natural extension of the
optical and near-IR studies of the sample already performed by our
group \citep[see][and references therein]{Heydari07,Meynadier07} which
were concentrated on the stellar content and excitation mechanisms of
the ionized gas. The present work is focused on the so far unexplored
mid-IR colors and morphology of these objects. This complete
multi-wavelength coverage should facilitate the comparison of the
object properties in various wavelengths, and enable a more realistic
modeling of these objects in the future. Given the direct coupling
between gas, dust, and star formation, the Spitzer data present new
possibilities in understanding the physics of obscured dust enshrouded
regions. It is thus instructive to contrast the compact and normal \h2
region samples in mid-IR.

We perform photometry using a circular aperture of $\sim$7\,\arcsec\
in diameter, which corresponds to a size of $\sim$ 2 pc at the
distance of the LMC and SMC.  The aperture size is based on the fact
that comparison between the mid-IR and optical images showed that the
sample objects are always less extended in the mid-IR band.  A total
of 27 blobs from the sample analyzed in \citet{Meynadier07} were in
areas of the Magellanic clouds where high quality Spitzer imaging data
were publicly available.  A particularity of this study is that it
deals with a sample of compact \h2 region which have comparable
physical sizes.  In Section 2 we present the sample and the Spitzer
observations, and in Section 3 we discuss our main results. We discuss
our findings in the context of similar studies of giant \h2 regions
and young dust enshrouded protostars in Section 4, and close with our
conclusion in Section 5.

\section{Observations and Data Reduction}

The data used for this study were obtained with the Infrared Array
Camera (IRAC) \citep{Fazio04} instrument on board the Spitzer Space
Telescope \citep{Werner04} and were recovered from Spitzer
Archive. The observations of the Large Magellanic Cloud (LMC) were
part of the Spitzer Legacy project ``Surveying the Agents of a
Galaxy's Evolution'' (SAGE, PI M. Meixner, PID = 20203, see Meixner et
al. 2006), while the Small Magellanic Cloud data were observed by the
GO program of A. Bolatto (PID = 3316, see Bolatto et al. 2007). A total
of 24 sec of on-source observing time per filter was devoted for each
sky position at the LMC using 12 sec frames and a raster which resulted
in two frames per position. The SMC data were obtained using 12sec
frames and a 3 position dither producing an on source exposure of a
total of 36 sec. The data were processed with the Spitzer Science
Center (SSC) pipeline (version 14.0.0) which created the final mosaics
used for our analysis.

All 27 sources were identified visually based on accurate coordinates
and after cross-checking their position with existing optical
observations. As an example of the quality of the data we present in
Fig. 1 an image of the general area in the LMC which contains the HEB
N11A.  Photometry was performed using the ``aper'' IDL routine with
circular apertures of 3 pixels (3.6$\arcsec$) in radius and the
corresponding flux correction factors for the extended size of the PSF
described in the IRAC data handbook were applied. Special care was
devoted to proper background subtraction as some of the sources were
in crowded fields. The absolute errors in photometry are at the 5\%
level. Our photometric results are presented in Table 1, which gives
the target ID used in Figs. 3 and 4, the RA and Dec of the target
identified in the IRAC images, and the Spitzer observation AORKEY. A
few sources of the original sample, presented in \citet{Meynadier07},
which were saturated or unresolved in the Spitzer images were not
included in this study.  In Table 1 we also provide the 1$\sigma$ of
the variation of the sky (in MJy sr$^{-1}$) next to each target.

\begin{figure*}[]
  \centering
  \includegraphics[width=1.0\linewidth]{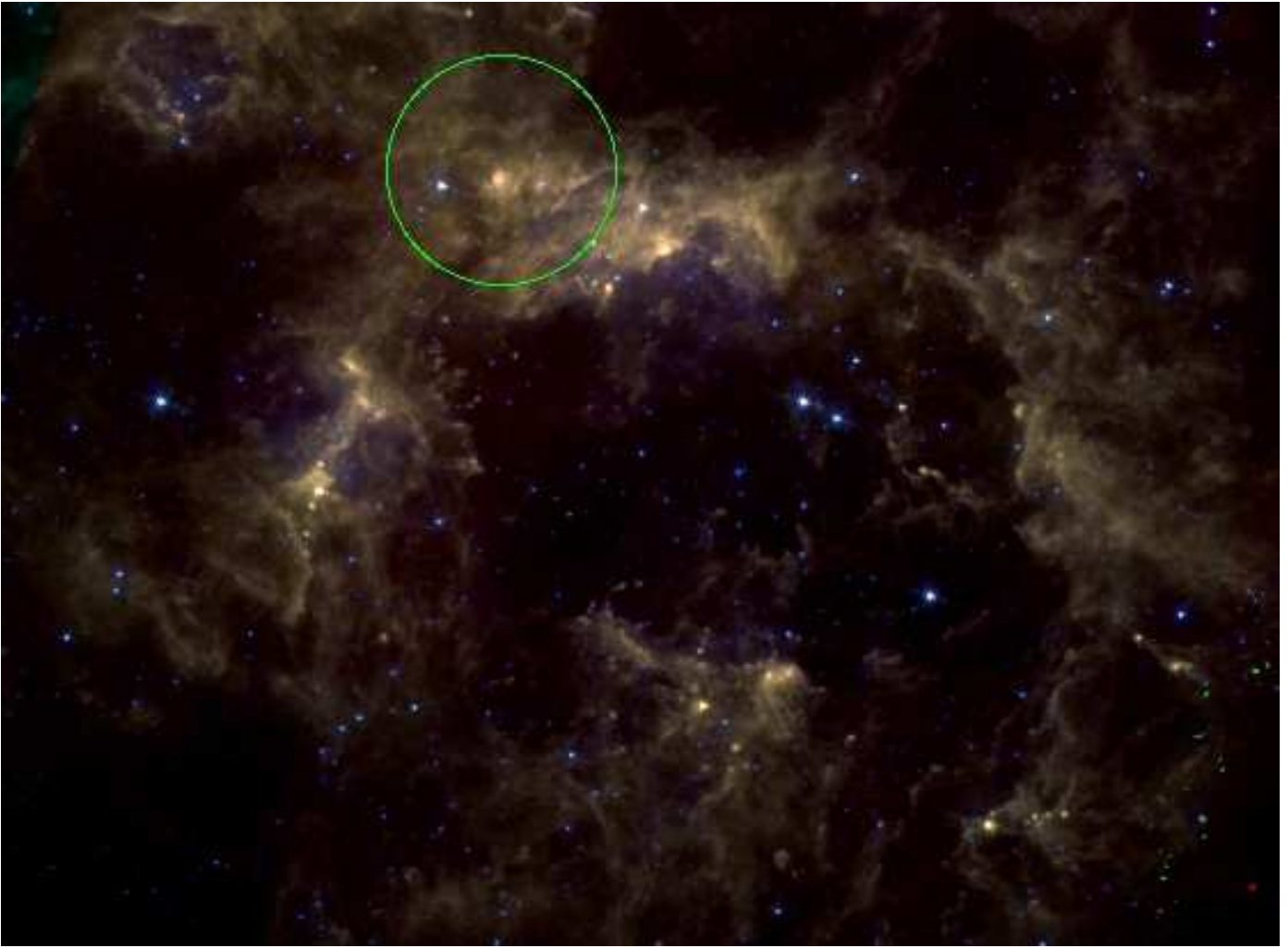}
  \caption{A ``true color'' image of the Large Magellanic Cloud giant
    \h2 region N11. A circle of 2 arcmin in radius is centered on the
    compact \h2 region N11A, which lies on the northern ridge of a
    giant shell surrounding a cavity created by massive stars in its
    central regions.  The composite image was created by combining the
    three 3.6, 5.8, and 8.0 $\mu$m bands into a blue-green-red
    frame. The image size is $\sim22.8\,\times$ 9.0 arcmin,
    corresponding to $\sim410\,\times$ 162 pc at the distance of the
    LMC. North is up and East is to the left. }
\end{figure*}

\begin{figure*}[]
  \centering
  \includegraphics[width=1.0\linewidth]{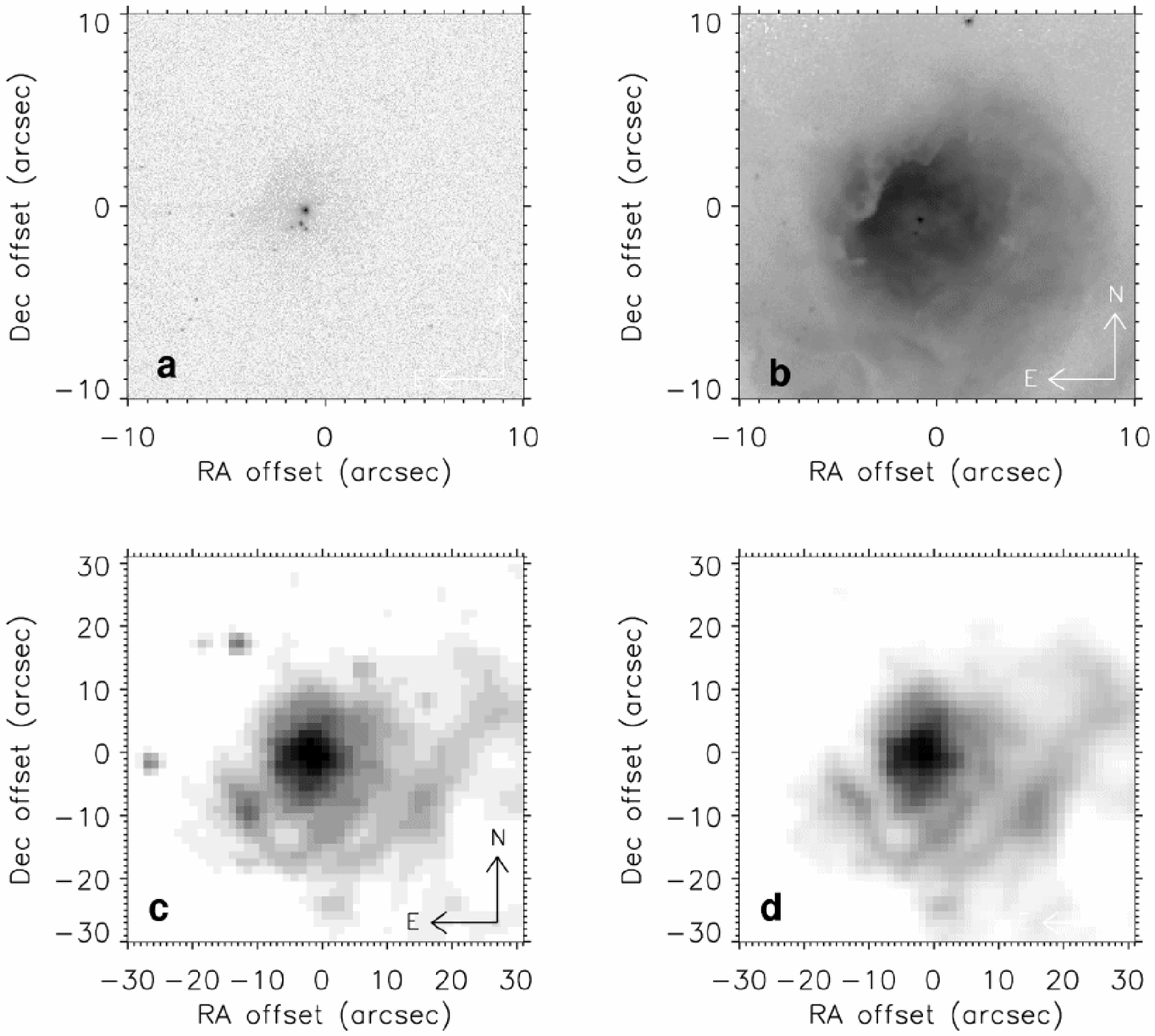}
  \caption{ A mosaic of four close-up images of the LMC
    high-excitation region N11A presented in Fig. 1. North is up,
    East is to the left, and the center of the field is at RA=4:57:16.2,
    Dec=-66:23:20.8 (J2000).  The upper-row images were obtained with {\it
      HST} \citep[see][]{Heydari01} and the lower-row ones with
    Spitzer. {\it a}) B-band image obtained with the F410M filter of
    WFPC2. A few of the central exciting stars are visible. {\it b})
    \ha\ image of N11A. Notice the turbulent environment surrounding
    the central source, the filamentary structure due to dust
    extinction and the ionization front the to the north west. {\it
      c}) A Spitzer/IRAC 3.6$\mu$m image of N11A. Even though extended
    emission is detected, the bulk of the light is well encompassed
    within our aperture. {\it d}) Same as in {\it c}) but for the
    Spitzer/IRAC 8$\mu$m image. }
\end{figure*}

\section{Results}

The analysis of the IRAC images of the regions resulted in
identification of point sources in the expected locations of the HEBs
and LEBs. Despite the improved Spitzer resolution, the size of the
IRAC point spread function (FWHM\,$\sim2.5\arcsec$ at 8 $\mu$m) does
not permit a direct comparison with the sub-arc-second resolution of
the {\it HST} optical images or the deconvolved VLT near-IR imagery
available \citep[i.e.][]{Heydari99,Heydari03}.  For example, Fig. 1
presents the LMC high-excitation blob N11A, which can be compared with
the {\it HST} image of this object \citep[see][for a detailed
analysis]{Heydari01}.  A series of close-up images of LMC-N11A is also
presented in Fig. 2. One can identify some of the exciting stars
visible in the {\it HST} B-band Fig. 2a image, as well as the
turbulent environment surrounding them, shaped by their strong stellar
winds, visible by the \ha\ emission displayed in Fig. 2b. In addition,
we include in Fig. 2c and Fig. 2d the Spitzer 3.6 and 8$\mu$m images
of the N11A. Even though we do detect some diffuse emission around the
region, more than 90\% of the flux originates from the central
``blob'', and it well sampled by our photometry aperture.  Thus, for
the purpose of this study, we shall consider all compact \h2 regions
as unresolved to this scale.

To explore the mid-IR properties of our sample we constructed the
typical IRAC magnitude and color-color diagrams presented in Fig. 3
and 4 and discussed in detail in the following section. In Fig. 3 we
display the 8 $\mu$m flux of the regions as a function of their
[3.6]-[8.0] color, with their ID, as indicated in Table 1. A box
marked with dashed lines indicates the expected locus of normal giant
\h2 regions at the mean distance of the Magellanic Clouds
\citep[][]{Whitney04, Meixner06, Gouliermis07}. In addition, the
general location where low-mass young stellar objects (YSOs), Class I
and Class II, have been found \citep[based on observations presented
by ][]{Cohen93, Cohen07, Gouliermis07} are also marked with the
hatched rectangular areas.  We observe that overall both HEBs and LEBs
are found in the areas of the diagram where normal extended \h2
regions and Class II protostars are located. This aspect will be
discussed in the following Section.  Furthermore, the HEBs are in
general more luminous than the LEBs and that the most luminous ones
are also the reddest. This is in line with previous results from
optical spectroscopy implying that HEBs are younger than LEBs
\citep{Meynadier07}. We note also that the variation of the mid-IR
luminosity at 8\,$\mu$m versus the [3.6] - [8.0] color can be roughly
approximated with a linear fit.

\begin{figure*}[]
  \centering
  \includegraphics[width=1.0\linewidth]{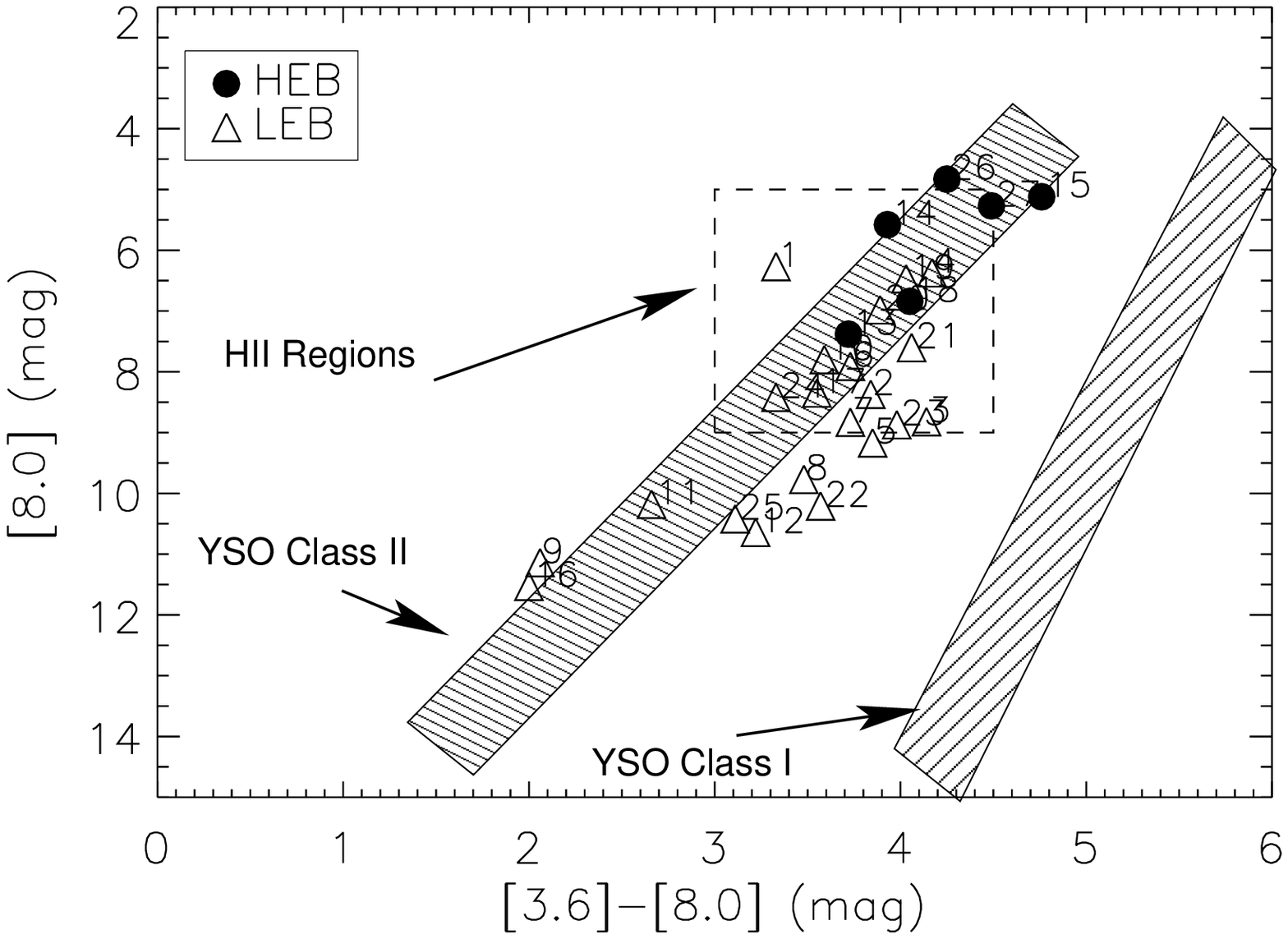}
  \caption{The {8.0 $\mu$m} flux (in mags) as a function of the
    [3.6]-[8.0] color.  The expected location of typical Galactic \h2
    regions as well as Class I and II low-mass YSOs is also included
    (see text). HEBs are marked with filled circles and LEBs with open
    triangles.}
\end{figure*}

\begin{figure*}
  \centering
  \includegraphics[width=1.0\hsize]{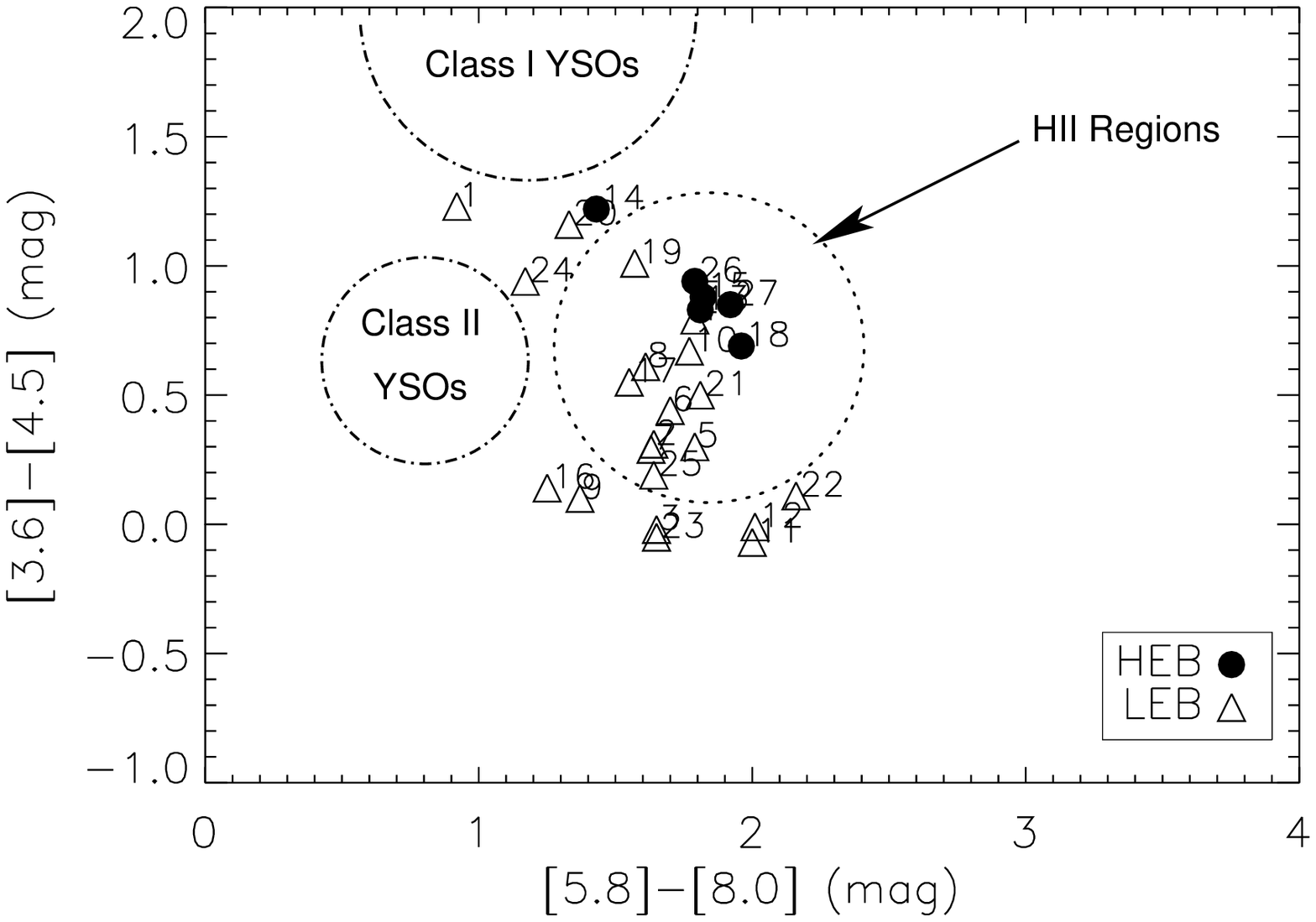}
  \caption{The [3.6]-[4.5] IRAC color as a function of the [5.8]-[8.0]
    color. We also indicate the expected positions for YSOs as well as
    the locus of giant \h2 regions from the models of
    \citet{Allen04}. HEBs are marked with filled circles and LEBs with
    open triangles.}
\end{figure*}

\section{Discussion}

Generally speaking, the mid-IR fluxes do not directly originate from
the exciting stars of the corresponding \h2 region, as photospheric
emission from individual OB stars at the distance of the Magellanic
Clouds is too faint in the [8.0] $\mu$m band to be measured
\citep{Jones05}.  The 8 $\mu$m IRAC filter probes mostly reprocessed
emission from the 7.7 $\mu$m PAH feature which is produced in the
photo-dissociation regions (PDRs) surrounding star forming regions
\citep[ie][]{Peeters02}.  The HEBs have higher 8 $\mu$m fluxes than
the LEBs because the former harbor hotter, more massive stars whose
ionizing radiation can penetrate deeper into the PDRs producing more
PAH feature emission.

Moreover, as pointed out previously, the HEBs are younger and
therefore are found in areas with larger quantities of gas and dust.
These two factors  contribute to the observed linear
relationship between the 8 $\mu$m flux and the [3.6] -- [8.0] color.
In addition to the elevated PAH emission from the PDRs falling within
our apertures, the 8 $\mu$m flux, along with the 4--12 $\mu$m slope
can also increase substantially due to emission from very small
grains. These grains have a radius less than 10 nm and are prominent
in Galactic \h2 regions \citep[]{Verstraete96,Cesarsky96} as well as
in deeply embedded extragalactic giant \h2 regions
\citep[i.e.][]{Mirabel98}.

Overall the LMC HEBs are more luminous and more dusty than their SMC
counterparts. Previous observations, using both ground-based
telescopes and {\it HST}, have shown N88A (object \#14) to be an
outstanding compact \h2 region in the SMC as far as its luminosity and
dust content are concerned. Fig. 3 confirms this showing this object
not only of a higher level than SMC N81 (object \#13), but also quite
comparable to the brightest blobs in the LMC.  Its dust content seems
to be stronger in the bands below 8 $\mu$m.  Among the LMC blobs, N11A
(\#18) is less luminous in the 8 $\mu$m band, while its mid-IR colors
indicate the presence of large quantities of dust, or increased PAH
emission.

It also appears that the average [3.6]-[8.0] color is bluer in LEBs 
suggesting a less dusty environment.  LEBs tend to be more evolved 
since their exciting stars have disrupted the bulk of their associated 
molecular clouds. Moreover, they are found in more isolated regions of 
the LMC and SMC, and therefore the \hi\ and CO column densities are on 
average lower. 

Fig. 1 is particularly interesting in that it shows N11A to be
situated in a giant mid-IR shell surrounding a cavity created by
energetic photons of the OB association LH9, which occupies the
central area of N11B \citep{Hatano06}. The association of N11A with
the shell is very likely not just a line of sight projection but there
seems to be a physical connection between the two. N11A may therefore
be a second generation object triggered by the ionization front due to
the expansion of N11B according to the scenario first suggested by
\cite{Elmegreen77}. This suggestion is in agreement with
\cite{Hatano06}, who find it quite likely that a new generation of
stars be formed in peripheral regions of N11B from swept-up molecular
clouds.  Examples of triggered massive-star formation at the border of
Galactic \h2 regions have recently been discussed in the literature
\citep[e.g.,][]{Deharveng05,Zavagno07}.

As we can see from Fig. 3, the colors of the sample compact \h2
regions are inconsistent with the locus of Class I YSO objects. There
is though some overlap with the Class II objects \citep[see also
][]{Gouliermis07}.  However, we find it quite unlikely that each HEB
or LEB in our sample be associated with a low-mass YSO. In their
study, using Spitzer IRAC observations, \citet{Jones05} detected a
couple of Class I YSO candidates towards the LMC giant \h2 region
N159, which spans over 5 arcmin on the sky, corresponding to 75
pc. Similarly, \citet{Gouliermis07} found 22 candidate YSOs in their
Spitzer study of the SMC \h2 region N90/NGC 602, which has a physical
extent of about 70 pc. In comparison, the physical sizes of our sample
blobs are around $\sim$2 pc. Consequently, the probability that each
\h2 blob harbors a low-mass YSO is very small, suggestion that the
overlap of the Class II YSOs with the blobs in Fig. 3 is fortuitous.

In order to get a better insight on the infrared properties of our
sample, we present in Fig. 4 the second diagnostic IRAC color-color
diagram where the [3.6]-[4.5] IRAC color is plotted as a function of
the [5.8]-[8.0] color. In this figure we mark again the expected loci
of giant \h2 regions and YSOs based on the models of \citet{Allen04}
dealing with Galactic low-mass YSOs \citep[see also][]{Jones05}.  We
note that the YSOs are in principle readily separated from the compact
\h2 regions.  Extinction from dust though will have as an effect to
move the data points of our sample towards the region occupied by
Class I YSOs. This could be the reason for the colors of SMC-N10
(object \#1). However, it seems unlikely that reddening be so
substantial for the majority of the sample. Based on the work of
\citet{Megeath04} an $A_{V}$ = 30\,mag would move the \h2 points by
0.5 mag towards the locus of Class I YSOs. Our past optical and
near-IR studies though suggest that the extinction towards the HEBs
and LEBs does not reach these high values.  The global mid-IR colors
for some blobs of our sample (i.e. objects \#3, \#9, \#23) are similar
to those found in PDR regions. This deduction is based on the \h2
region results of \citep[][]{Zavagno06} where they find that towards a
pure filament devoid of any exciting star, the mid-IR colors are
[3.6]-[4.5]\,$\sim$\,0.1 mag and [5.8]-[8.0]\,$\sim$\,1.8
mag. Elevated PAH emission can contribute to a red [5.8]-[8.0] color
while photospheric emission from stars results to a zero or negative
[3.6]-[4.5] color.

It is interesting to note that even though all compact \h2 regions in
our sample have approximately the same size in the optical, their
H$\alpha$ flux varies by nearly two orders of magnitude, with the HEBs
being the more luminous \citep{Meynadier07}. Consequently their 8
$\mu$m emission, which has been shown to be correlated in general with
to the \ha\ emission \citep[][]{Wu05, Calzetti07}, also spans over the
same range, nearly 6 mags (see Fig. 3).  Using the \ha\ and the
de-reddened \hb\ photometry of \citet{Meynadier07}, we searched for
such a correlation for our sample. However, we found that these ratios
exhibit a substantial scatter which is similar in both LEBs and
HEBs. It is conceivable that this could be due to the small size of
the compact \h2\ regions and the fact that our Spitzer apertures do
contain some mid-IR emission from the surrounding regions, as well as
to uncertainties in the \hb\ extinction corrections.

\section{Conclusions}

Using Spitzer archival data we have explored the mid-IR properties for
a sample of 27 high- and low-excitation compact \h2 regions in the
Magellanic Clouds. This first study of the blobs in mid-IR is a
necessary step towards understanding various aspects of these objects.

We find that in spite of their similar linear sizes (of just $\sim$2
pc) the mid-IR colors of these objects are comparable to typical giant
Magellanic Cloud \h2 regions. HEBs appear more luminous than LEBs at 8
$\mu$m, which is consistent with their higher \ha\ and \hb\
emission. HEBs are also on average redder since they are younger and
enshrouded in larger quantities of dust. No variations in the mid-IR
colors were found to be correlated to metallicity or hardness of the
radiation field of the sources. Although the loci of the blobs in the
mid-IR color-magnitude plot overlap with low-mass YSOs of Class II,
the probability that they contain such an object is estimated to be
very low.

\begin{acknowledgements}
  We would like to thank Dr. Lise Deharveng, Laboratoire
  d'Astrophysique de Marseille, for stimulating comments and
  discussions. VC would like to acknowledge partial support from the
  EU ToK grant 39965.

\end{acknowledgements}

\begin{table*}
\caption{Mid-IR photometry of the Compact H II Region Sample}             
\label{table:1}      
\centering          
\begin{tabular}{c c c c c r r r r r r r r}     
\hline\hline       
ID & Region & RA (J2000) & Dec (J2000) & AORKEY & 
\multicolumn{4}{c}{Source (mag)} & \multicolumn{4}{c}{Sky (MJy sr$^{-1}$)}\\
\hline

 & &  & &  & 
f$_{3.6\mu m}$  & f$_{4.5\mu m}$  & f$_{5.8\mu m}$  & f$_{8.0\mu m}$ &
$\sigma_{3.6\mu m}$  & $\sigma_{4.5\mu m}$  & $\sigma_{5.8\mu m}$  & $\sigma_{8.0\mu m}$ \\  
\hline                    

1    &   \object{SMC-N10}	&0:44:56.4 &   -73:10:10.7	&10740992   &    9.60	&8.37	&7.19	&6.27	&0.159	&0.178	&0.369	&0.569	\\
2    &   \object{SMC-N11}	&0:45:00.2 &   -73:16:37.4	&10740992   &    12.21	&11.90	&10.01	&8.37	&0.515	&0.452	&0.642	&0.834	\\
3    &   \object{SMC-N21}	&0:47:53.1 &   -73:17:35.2	&10740992   &    12.96	&12.98	&10.47	&8.82	&0.287	&0.352	&0.760	&0.890	\\
4    &   \object{SMC-N26}	&0:48:08.6 &   -73:14:55.6	&10740992   &    10.54	&9.75	&8.15	&6.37	&0.399	&0.493	&0.762	&1.084	\\
5    &   \object{SMC-N31}	&0:48:41.9 &   -73:26:15.6	&10740992   &    13.03	&12.73	&10.96	&9.17	&0.419	&0.552	&0.324	&0.836	\\
6    &   \object{SMC-N33}	&0:49:28.9 &   -73:26:34.2	&10743040   &    11.64	&11.20	&9.61	&7.91	&0.476	&0.278	&0.193	&0.193	\\
7    &   \object{SMC-N32}	&0:49:40.9 &   -72:48:47.9	&10741248   &    12.56	&12.27	&10.45	&8.83	&0.778	&0.852	&0.358	&0.708	\\
8    &   \object{SMC-N45}	&0:51:40.7 &   -73:13:34.6	&10743040   &    11.02	&11.14	&11.03	&10.86	&0.123	&0.310	&0.313	&0.484	\\
9    &   \object{SMC-N64}	&0:58:16.8 &   -72:38:38.5	&10741504   &    13.20	&13.09	&12.51	&11.14	&0.238	&0.328	&0.276	&0.499	\\
10   &   \object{SMC-N68}	&0:58:42.9 &   -72:27:16.8	&10741504   &    11.39	&10.72	&9.57	&7.80	&0.156	&0.326	&0.211	&0.262	\\
11   &   \object{SMC-N75}	&1:02:31.6 &   -71:57:00.3	&10742016   &    12.84	&12.90	&12.18	&10.18	&0.111	&0.766	&0.676	&0.239	\\
12   &   \object{SMC-N77}	&1:02:49.2 &   -71:53:19.0	&10742016   &    13.86	&13.87	&12.65	&10.64	&0.335	&0.115	&0.151	&0.489	\\
13   &	 \object{SMC-N81}$^*$	&1:09:13.0 &   -73:11:38.3	&10742528   &    11.10	&10.27	&9.19	&7.38	&0.091	&0.652	&1.049	&1.466	\\
14   &	 \object{SMC-N88A}$^*$	&1:24:08.0 &   -73:09:03.9	&4383488    &9.51	&8.29	&7.01	&5.58	&0.040 &0.028	&0.086	&0.123	\\
\hline																					
15   &   \object{LMC-N83B}$^*$	&4:54:25.2 &   -69:11:03.2	&14351616  &9.88	&9.00	&6.94	&5.12	&0.152	&0.274	&0.271	&0.331	\\
16   &   \object{LMC-N88}	&4:54:51.6 &   -69:23:24.4	&14353152   &    13.53	&13.39	&12.77	&11.53	&0.308	&0.109	&0.142	&0.151	\\
17   &   \object{LMC-N90}	&4:55:25.5 &   -69:16:06.4	&14353152   &    11.89	&11.35	&9.90	&8.35	&0.509	&0.091	&0.501	&0.170	\\

18 & 	 \object{LMC-N11A}$^*$	&4:57:16.2 &   -66:23:20.8	&14352128  &10.89	&10.20	&8.79	&6.84	&1.409	&1.010	&1.331	&2.478	\\
19   &   \object{LMC-N191A}	&5:04:37.5 &   -70:54:37.1	&14354944   &    10.50	&9.48	&8.03	&6.47	&6.011	&0.512	&4.648	&1.546	\\
20   &   \object{LMC-N105A}	&5:09:48.6 &   -68:52:44.0	&14353664   &    10.87	&9.71	&8.31	&6.98	&0.473	&0.508	&0.907	&0.955	\\
21   &   \object{LMC-N193}	&5:12:32.2 &   -70:24:19.3	&14367488   &    11.66	&11.16	&9.40	&7.60	&0.097	&0.123	&0.166	&0.247	\\
22   &   \object{LMC-N33}	&5:16:47.6 &   -67:19:38.4	&14366464   &    13.79	&13.68	&12.38	&10.21	&0.088	&0.148	&0.357	&0.454	\\
23   &   \object{LMC-N197}	&5:20:53.9 &   -71:43:16.1	&14356736   &    12.85	&12.90	&10.52	&8.87	&0.780	&0.073	&0.572	&0.266	\\
24   &   \object{LMC-N68}	&5:37:02.8 &   -68:14:11.0	&14357504   &    11.76	&10.81	&9.59	&8.42	&0.120	&0.053	&0.765	&0.574	\\
25   &   \object{LMC-N156}	&5:37:38.2 &   -69:34:25.4	&14371584   &    13.53	&13.34	&12.06	&10.42	&0.418	&0.342	&0.311	&0.957	\\
26 & 	\object{LMC-N160A2}$^*$	&5:39:46.1 &   -69:38:35.5	&14371584  &9.08	&8.14	&6.62	&4.83	&0.077	&0.221	&0.957	&1.931\\
27 &	\object{LMC-N159-5}$^*$	&5:40:04.4 &   -69:44:37.4	&14359040  &9.76	&8.91	&7.19	&5.27	&0.651	&0.520	&1.120	&1.464	\\

\hline       
\\           
\end{tabular}
{\bf Table note:} The horizontal line separates the SMC and LMC
  sources. The high excitation blobs (HEBs) of our sample are
  indicated with an asterisk ($^*$).

\end{table*}


\end{document}